\documentclass[12pt]{iopart}
\usepackage{graphicx,cite}
\begin{document}
\article{\letter{}}{Point perturbations of circle billiards}

\author{S Rahav\dag, O Richman\ddag, and S Fishman\dag}
\address{\dag\ Department of Physics, Technion, Haifa 32000, Israel}
\address{\ddag\ Department of Electrical Engineering, Technion, Haifa 32000, Israel}

\begin{abstract}
The spectral statistics of the circular billiard with a point-scatterer
is investigated. In the semiclassical limit, the spectrum is
demonstrated to be composed of two uncorrelated level sequences.
The first corresponds to states for which the scatterer
is located in the classically forbidden region and its energy levels
are not affected by the scatterer in the semiclassical limit while
the second sequence contains the levels which are affected by the point-scatterer.
The nearest neighbor spacing distribution which results from the superposition
of these sequences is calculated analytically within some approximation
and good agreement with the distribution that was computed 
numerically is found.
\end{abstract}

\nosections
Classical dynamics may be illuminating for the understanding of the
corresponding quantum mechanical systems.
One of the most studied aspects of the relation
between classical and quantum mechanics is the connection between
the spectral statistics of the quantum system and the dynamical 
properties of its classical counterpart.
Classically integrable systems typically exhibit Poisson-like
spectral statistics~\cite{BT77} while classically chaotic
systems exhibit spectral statistics of random matrix
ensembles~\cite{BGS,berry85,bohigas,mehta}. The spectral statistics 
of integrable and chaotic systems are universal, that is, 
they do not depend on specific details of the system but
rather on the type of motion and its symmetries.
There are systems which are intermediate between integrable and chaotic
ones and their spectral properties are not known to be
universal. Such systems are of experimental relevance. The spectral
statistics of mixed systems, for which the phase space is
composed of both integrable and chaotic regions,
were studied by Berry and Robnik~\cite{BR84}. The spectrum can be viewed 
as a superposition of uncorrelated level sequences, 
corresponding to the various regions,
which are either chaotic or integrable.
The nearest neighbor spacing distribution (NNSD) of such a superposition
of sequences was calculated in~\cite{BR84}.
The resulting statistics are, in some sense,
intermediate between those of integrable and chaotic systems.
Other types of systems with intermediate statistics
include pseudointegrable systems
and integrable systems with flux lines or point-scatterers~\cite{rahav01,bogomolny01,bogomolny01b,seba90,rahav02,bogomolny02,bogomolny02b}.

The spectral statistics of billiards with flux lines~\cite{rahav01,bogomolny01}
and of some pseudointegrable billiards~\cite{bogomolny01} were recently studied.
A possible route towards an understanding of the spectral statistics 
of these systems
is based on their classical periodic orbits. It is possible to compute the
correlation function of the energy levels from these orbits by using trace 
formulae~\cite{gutz,BT}.
For the billiards with flux lines or for pseudointegrable billiards,
the orbits include not only the periodic
orbits but also diffracting orbits which are built from segments that
start and end at some singularity of the system. While the contributions
from periodic orbits were easily calculated, those from diffracting orbits turn
out to be much more involved.
The spectral statistics of these systems, that can be obtained
numerically, appear to be intermediate
between those of integrable and chaotic systems. In particular, the NNSD show level repulsion
at small spacings and an exponential falloff at large spacings.

Since the contributions of diffracting orbits to the spectral statistics of
pseudointegrable systems and of billiards with flux lines is far from being understood, it is
of interest to study simpler systems which exhibit intermediate statistics.
A class of such systems is given by integrable systems with a point-scatterers.
A point-scatterer is the self adjoint extension of a ``$\delta$-function potential''
in two or three dimensions~\cite{pointbook}.
The spectral statistics of an integrable system with such a point-scatterer
was first studied by \v{S}eba~\cite{seba90}.
It is a rectangular billiard with the
perturbation at its center. This ``\v{S}eba billiard'' also exhibits 
intermediate statistics which differs from that of psedointegrable systems.
Integrable systems with point-scatterers are much easier to study analytically
compared to integrable systems with flux lines or to pseudointegrable systems.
The contributions of diffracting orbits to the correlation function of the energy levels were 
recently calculated for the rectangular billiard with a point-scatterer~\cite{rahav02,bogomolny02}.
Exact results for the NNSD were also obtained \cite{bogomolny02b}.
One of the intriguing features of the spectral statistics
of the rectangular billiard with a point-scatterer (and Dirichlet boundary
conditions) is its dependence on the location of the scatterer.
If the coordinates of the scatterer (divided by the sides of the rectangle)
are rational numbers $p_i/q_i$ (where $i=x,y$) then the spectral statistics depend in a non trivial way
on $p_i,q_i$ \cite{bogomolny02b}.
In contrast, for typical locations, the spectral statistics seem to be location independent.
The cause for this dependence on location is that many wavefunctions vanish at rational values of the 
coordinates. At these locations there are many
degeneracies in the lengths of the diffracting orbits (including
repetitions). Since such dependence on location is 
atypical, it is of interest to study the dependence of the spectral statistics
on the location of the perturbation in other systems. For instance, it may be possible
that for typical systems the location of the point scatterer affects the spectral
statistics only smoothly (and does not depend on the rationality of the
coordinates of the scatterer). 

The system that is studied in this work is the circle billiard 
perturbed by a point-scatterer and
the dependence of the spectral statistics on its location is 
studied. The (two dimensional) circle billiard with radius $R$ is described by the Schr\"odinger equation
\begin{equation}
\label{schr}
-{\Delta} \psi = E \psi
\end{equation}
with Dirichlet boundary conditions $\psi(|{\bf r}|=R)=0$. 
(The units where $\hbar=2m=1$ are used in most of this work.)
The Hamiltonian with the point-scatterer
is the self-adjoint extension of a Hamiltonian where one point, say ${\bf x}_0$, is
removed from its domain. It can be considered as the self-adjoint extension of a Hamiltonian
with a $\delta$-function potential at ${\bf x}_0$.
Given the eigenvalues (and eigenfunctions) of the unperturbed system $E_n$ (and $\psi_n$), namely the system in absence of the $\delta$-scatterer,
the eigenvalues of the system with the point-scatterer are given by
the roots of~\cite{zorbas80,seba90}
\begin{equation}
\label{solvept}
\fl \left( \frac{\sin \xi}{1-\cos \xi} \right) \sum_n \left| \psi_n ({\bf x}_0) \right|^2 \frac{\Lambda}{\Lambda^2 + E_n^2} - \sum_n \left| \psi_n ({\bf x}_0) \right|^2 \left( \frac{1}{z-E_n} + \frac{E_n}{E_n^2+\Lambda^2} \right) = 0,
\end{equation}
where $\Lambda$ and $\xi$ are two parameters. For a more complete discussion regarding this equation, the roles of the parameters as well as the method of its numerical solution see, for example,~\cite{rahav02}.
This equation turns out to be very convenient for numerical solution since every root is located between
two eigenvalues of the unperturbed system.

The (unnormalized) eigenfunctions of the circle billiard are
\begin{equation}
\label{functions1}
\tilde{\psi}_{n,m}=e^{\pm i m \phi} {\cal J}_m (k_{n,m} r)
\end{equation}
where ${\cal J}_m$ are Bessel functions of the first kind
and the angular momentum $m$ is any nonnegative integer. The 
energy levels $E_{n,m}=k_{n,m}^2$ are determined by the
boundary condition ${\cal J}_m (k_{n,m}R)=0$. 
It is obvious that all the energy levels with $m \ne 0$ are doubly degenerate. 
As a result there is a linear combination of the two degenerate 
wavefunctions which vanishes at ${\bf x}_0$ and an orthogonal linear combination 
which does not vanish at ${\bf x}_0$.
The perturbation breaks this degeneracy.
The linear combination which vanishes is also an eigenfunction of
the Hamiltonian {\em with} the point-scatterer and thus $E_{n,m}$
is an eigenvalue of the perturbed problem. Therefore, half of the spectrum
is unchanged by the perturbation. To avoid this trivial part of
the spectrum we choose to work with the non vanishing
linear combinations (which are eigenfunctions of the unperturbed Hamiltonian)
and only the half of the spectrum which is affected by the perturbation
will be considered in this work. 
For convenience, the location of the perturbation
is chosen at ${\bf x}_0=(r_0,0)$ and therefore the eigenfunctions of the
unperturbed Hamiltonian which do
not vanish there are
\begin{equation}
\label{functions2}
\psi_{n,m} (\phi,r)= \sqrt{2/\pi(1+\delta_{m,0})} \left( R {\cal J}_{m+1} ( k_{n,m} R)\right)^{-1} \cos m \phi \; {\cal J}_m (k_{n,m} r).
\end{equation}
The spectrum is determined by substituting these eigenfunctions 
and the corresponding energies $E_{n,m}=k_{n,m}^2$ in equation (\ref{solvept}).

We are interested in the dependence of the spectral statistics on
the location of the scatterer. This dependence can be easily understood
in terms of the properties of the wavefunctions of the unperturbed
system.
The quantum numbers $n,m$ correspond to a state with an angular momentum
$L=\hbar m$ and an energy $E_{n,m}=\hbar^2 k_{n,m}^2 / 2m$.
In the semiclassical limit where $\hbar \rightarrow 0$,
but $E$ and $L$ are kept fixed, the values of the wavefunctions are small in
the classically forbidden region $r<r_{min}=L / \sqrt{2 m E}$.
Therefore, one can divide the states into two groups.
The first consists of states for which the point-scatterer is located in the
classically forbidden region. As will be demonstrated, the
eigenvalues of these states change only slightly due to the perturbation
(and do not change at all in the semiclassical limit).
The second group includes the states for which the perturbation is
located in the classically allowed region. These states will be (strongly)
affected by the perturbation. This separation of the spectrum into
a superposition of two sequences is the cause for the dependence
of the spectral statistics 
on the location of the scatterer. This separation into
(semiclassically) affected and unaffected states is justified in the following.

To demonstrate that some of the eigenvalues are almost unchanged
by the perturbation one should solve equation (\ref{solvept}) and show that
for (exponentially) small eigenfunctions the corresponding
eigenvalues are almost unaffected. For simplicity, instead of 
equation (\ref{solvept})
it is sufficient
to consider the finite sum 
\begin{equation}
\label{finitept}
\sum_{i=1}^{N} \frac{|\psi_i|^2}{z-E_i}= a(z)
\end{equation}
where $a(z)$ is assumed to be slowly varying function of $z$.
We denote $\psi_i ( {\bf x}_0) \equiv \psi_i$.
To further simplify the argument let us assume that $|\psi_l|^2=A$
and $|\psi_{m}|^2=B$ are of order unity while for $i \ne l,m$
the wavefunctions on the scatterer $|\psi_i|^2 \equiv \epsilon_i$ are all small. The solutions of (\ref{finitept})
are to a good approximation given by the solutions of
\begin{equation}
\label{approx1}
\frac{A}{z_0-E_l}+\frac{B}{z_0-E_m} = a (z_0)
\end{equation}
for $z_0$ and by
\begin{equation}
\label{approx2}
z_i=E_i
\end{equation}
for $i \ne l,m$.
This is true since substituting a solution of the form 
$\tilde{z}_0 = z_0 + \delta z_0$ in (\ref{finitept})
leads to
\begin{equation}
\label{z0deltaz0}
\frac{A}{z_0+\delta z_0-E_l}+ \mathop{\sum_{i=1}^{N}}_{i \ne l,m} \frac{\epsilon_i}{z_0+\delta z_0 - E_i}+\frac{B}{z_0+\delta z_0 - E_m} = a (z_0 + \delta z_0).
\end{equation}
Expanding with respect to $\delta z_0$ and then solving to the leading 
order in $\epsilon_i$ results in
\begin{equation}
\label{correction1}
\delta z_0 \simeq 
\left(\frac{A}{(z_0-E_l)^2}+\frac{B}{(z_0-E_{m})^2}+a^\prime (z_0) \right)^{-1} \mathop{\sum_{i=1}^{N}}_{i \ne l,m} \frac{\epsilon_i}{z_0-E_i}.
\end{equation}
When $\epsilon_i \rightarrow 0$, as is the case in the semiclassical limit, $\delta z_0$
also approaches $0$.
For the other eigenenergies one can substitute $z_i=E_i + \delta z_i$
(with $i \ne l,m$)
and find that in the leading order
\begin{equation}
\label{correction2}
\delta z_i = \epsilon_i \left({a(E_i)-\frac{A}{E_i-E_l}-\frac{B}{E_i-E_{m}}}\right)^{-1}
\end{equation}
which also vanish when $\epsilon_i \rightarrow 0$.
Note that we have just found $N-1$ (approximate) solutions
which are all the solutions between $E_1$ to $E_N$.
It is not hard to generalize this calculation for more wavefunctions
which are of order unity. 
In the semiclassical limit,
the resulting spectrum
always consist of such affected and unaffected components.
In this limit the values
of the wavefunctions at the scatterer
are exponentially small if it is located in the classically forbidden
 region and the corresponding eigenvalues can be treated as unchanged
by the perturbation for any semiclassical consideration.

\begin{figure}
\centering
\includegraphics[width=12cm,height=10cm]{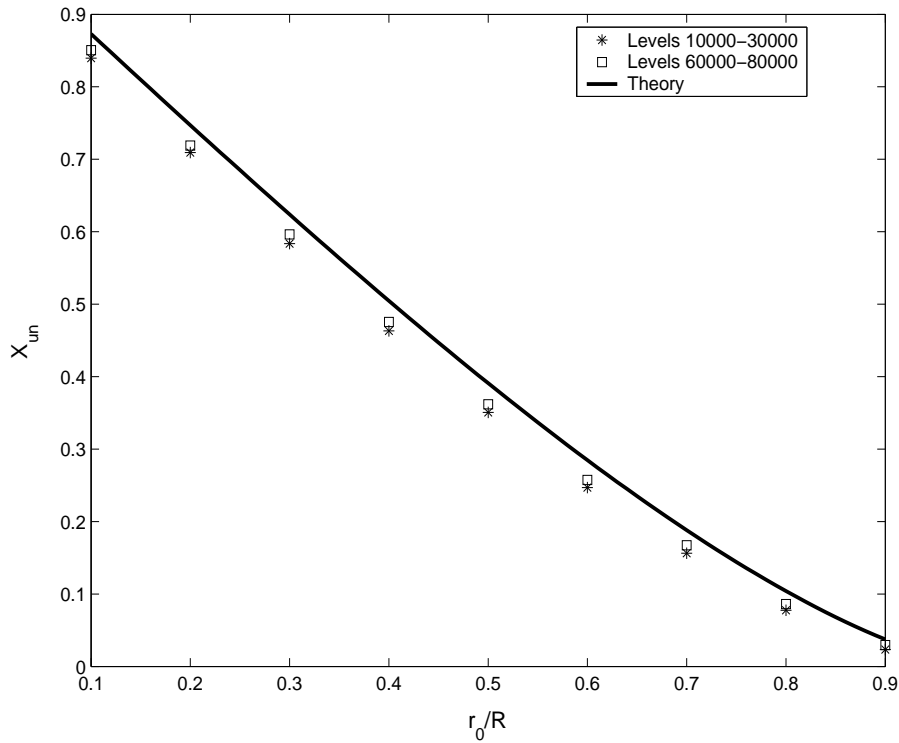}
\caption{The fraction of unaffected energy levels for two energy windows (stars - levels 10000 to 30000; squares - levels 60000 to 80000) compared to the theoretical prediction of Eq. (\protect{\ref{xunaff}}) (solid line). \label{unfig}}
\end{figure}

Consider a circle billiard with the point-scatterer at $r=r_0$. The 
spectral statistics of its energy levels in an energy window of width
$\Delta E$ around $E_0$ are studied in what follows.
Assume that $E_0$ is very large compared to $\Delta E$, 
i.e. that all the levels in the window have similar energies
which are high enough to be considered semiclassical.
A natural question to ask is how many of these levels are
affected by the point-scatterer and how many are unaffected
by it. As was argued, the semiclassically unaffected levels are those for which
the classical turning point, $r_{min}$, satisfies $r_{min}>r_0$.
Equivalently, for a given energy $E_0$,
for a state to be unaffected, the angular momentum
$L$ should satisfy $L>L_0=r_0 \sqrt{2 m E_0}$ 
(for the estimate of $L_0$, the values of the energies are approximated
by $E_0$).
The fraction of such states was calculated in~\cite{rahav99}
where it was used to determine how many levels are poorly approximated 
in the WKB method. This fraction is
\begin{equation}
\label{xunaff}
X_{un}=\frac{2}{\pi} \left[ \cos^{-1} \left( \frac{r_0}{R} \right)-  \frac{r_0}{R} \sqrt{1-\frac{r_0}{R}} \right].
\end{equation} 
It is clear that when $r_0 \rightarrow 0$ then $X_{un} \rightarrow 1$,
as expected, while for $r_0 \rightarrow R$, $X_{un} \rightarrow 0$.
As one approaches the semiclassical limit
these states are less and less affected. The predictions
of equation (\ref{xunaff}) can be checked numerically. The number
of unaffected levels was calculated for two energy windows
as a function of the location of the perturbation.
The results are presented in figure \ref{unfig}.
The radius of the billiard was
chosen so that the mean level spacing is $1$. Therefore both energy windows
contain $20000$ levels. A level was counted as unaffected
if its difference from an energy level of the unperturbed system was
less than $10^{-4}$ of the mean level spacing.
This criterion is somewhat arbitrary, since in
the semiclassical limit the difference can be taken to be arbitrarily
small.
 Figure \ref{unfig} indicates that equation (\ref{xunaff})
correctly describes the number of unaffected levels.
There is a slight deviation which is smaller for the levels from the
 higher energy window. This deviation is caused by the fact that 
for any finite energy the 
wavefunctions do not vanish
at the turning point $r_{min}$ but rather exhibit an Airy-like structure
(in the radial direction) near the turning point.
This means that for states with $r_{min}>r_0$, for which $r_{min}$ is
close to $r_0$, $|\psi ({\bf x}_0)|^2$ might not be small at a finite (but large) energy. 
The deviations are expected to
vanish in the semiclassical limit as indicated by figure \ref{unfig}.

The spectrum of the circle billiard with a point-scatterer can therefore be viewed
as composed of two uncorrelated components. One is unaffected by the point-scatterer
and its relative fraction is $X_{un}$ while the other is affected and its
relative fraction is $1-X_{un}$.
The NNSD of a spectrum which is composed of several uncorrelated level sequences
was computed by Berry and Robnik~\cite{BR84}
and is applied to the circle billiard with the point-scatterer
 in what follows. 

The unaffected spectrum consists of many levels with different
angular momentum quantum numbers and thus its statistics are Poissonian~\cite{BT77,BR84}.
Since the density of levels in this sequence is $X_{un}$, and the radius was chosen so that the total level density is unity, its NNSD
is 
\begin{equation}
\label{unaffps}
P_1 (S) = X_{un} e^{-X_{un} S}.
\end{equation}
The second level sequence contains the levels which are influenced by the
point-scatterer and their density is $1-X_{un}$.
The exact form of its NNSD is unknown and
an exact computation of this NNSD is complicated and beyond of the scope of this letter.
Instead, following experience with other systems~\cite{bogomolny01b,wiersig03},
we will {\em assume} that the NNSD can be {\em approximated} by a semi-Poisson distribution,
that is,
\begin{equation}
\label{affps}
P_2 (S) = 4 (1-X_{un})^2 S e^{-2 (1-X_{un})S}.
\end{equation}
This distribution exhibits level repulsion at small spacings and exponentially
small probability to find large spacings. In these works it was found numerically to describe
the distribution of spacings reasonably well but
there is no analytical justification for its use.
Note that even if the semi-Poisson distribution is an approximation for
the NNSD it may not approximate other spectral measures well.
For instance, the form factor, which is the Fourier transform of the 
energy-energy correlation function, satisfies $K(0)=1/2$
for the semi-Poisson distribution~\cite{rahav01} while for a billiard with point-scatterer one expects to find $K(0)=1$~\cite{rahav02}. In particular,
for the rectangular billiard with a point-scatterer, the NNSD was computed
analytically under some assumptions
in~\cite{bogomolny02b} and was found {\em not} to be given by
the semi-Poisson or the Poisson distributions. The nth neighbor spacing distributions were
also calculated there and found to be those of the Poisson distribution at
large spacings.

\begin{figure}
\centering
\includegraphics[width=12cm,height=10cm]{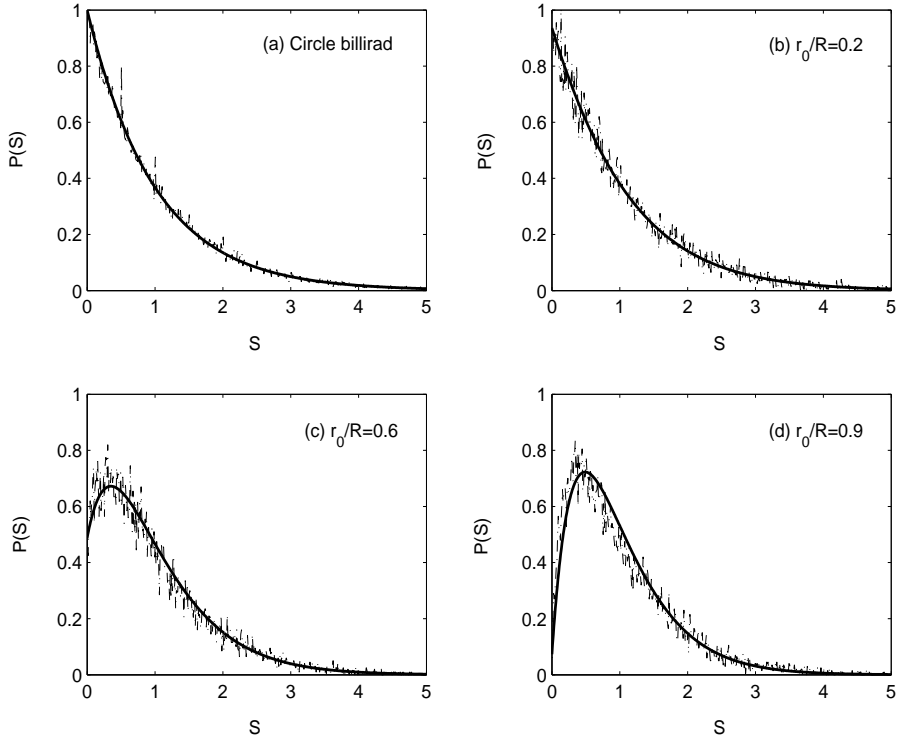}
\caption{The theoretical nearest neighbor spacing distribution (solid line),
given by  equation (\protect{\ref{totalps}}),
compared to numerical results (dashed-dotted line) for: (a) the circle billiard in absence of the scatterer, and with a scatterer at (b) $r_0/R=0.2$, (c) $r_0/R=0.6$ and (d) $r_0/R=0.9$. \label{pofs}}
\end{figure}

Following \cite{BR84} the NNSD of the circle billiard with the point-scatterer, obtained
by superposing the two sequences, is given by
\begin{equation}
\label{totalps}
P(S)=\left[X_{un} (2-X_{un}) + (1-X_{un})(2-X_{un})^2 S \right] e^{-(2-X_{un})S}.
\end{equation}
When the perturbation is at the center, $X_{un} \rightarrow 1$, and $P(S)$ approaches
the Poisson distribution. Alternatively, when the perturbation is near the boundary,
and $X_{un} \rightarrow 0$, the NNSD approaches the semi-Poisson distribution.
Note that the distribution (\ref{totalps})  does not exhibit complete level
repulsion since its value at $S=0$, $P(0)=X_{un} (2-X_{un})$, does
not vanish and it is a manifestation of the existence of
an infinite class of states that are unaffected by the
perturbation.
The NNSD of (\ref{totalps}) is compared to numerical results in figure~\ref{pofs}.
The NNSD was computed using the levels $40000$-$60000$ for three locations of the 
point-scatterer as well as for the circle billiard without the perturbation. It is clear
that the agreement is very good. The main features of the distribution are captured by
the simple argument leading to (\ref{totalps}). There are slight deviations from
the predictions of equation (\ref{totalps}), mainly at large $r_0/R$. These can be
attributed to the fact that the NNSD of the affected spectrum differs from the
semi-Poisson distribution.

The results presented in figures \ref{unfig} and \ref{pofs} suggest that the spectrum of 
the circle
with a point-scatterer consists of a superposition of two 
uncorrelated level sequences.
The relative densities of these sequences are determined by the way the classical tori
of the integrable system are projected into coordinate space. States for which the
perturbation is in the classically allowed region are affected while states for which the
perturbation is in the classically forbidden region are nearly unaffected. 
We expect this behavior to be typical of systems where the
scatterer affects only a fraction of the tori of the otherwise
classically integrable systems. This differs from the rectangular
billiard where all tori are affected by the scatterer.
Another important difference compared to the rectangle billiard
results of the different nature of the wavefunctions. For the
rectangular billiard there are infinite classes of wavefunctions
that have common zeros at rational points. Consequently, if the
scatterer is placed at such a location the wavefunctions are
not affected, and the distribution depends strongly on the 
rationality of the location of the scatterer. For the circle
billiard studied in the present work, on the other hand, there
is one class of eigenfunctions that vanish on the scatterer
since they are antisymmetric in $\phi$. These are not considered
in the present work.
The symmetric eigenfunctions always satisfy $\cos m \phi_0 =1$.
To obtain many functions, symmetric in $\phi$, that vanish at the same location
is equivalent to finding many Bessel functions
which satisfy ${\cal J}_m (k R) = 0 = {\cal J}_{m+l} ( k_1 R)$ and
${\cal J}_m (\alpha kR) = 0 = {\cal J}_{m+l} (\alpha k_1 R)$
for integer $m$, $l>0$ and for $\alpha=r_0/R<1$.
Finding an infinite number of such solutions, for the same $\alpha$
(corresponding to the same location of the perturbation),
is unlikely.
However, since any Bessel function of large argument is asymptotically
given by a cosine one can find states with close zeros,
that is where $k R$ and $\alpha k R$ are zeros of ${\cal J}_m$, 
while $k_1 R$ is a zero of ${\cal J}_{m+l}$ and $\alpha k_1 R$ is close 
to a zero of ${\cal J}_{m+l}$.
In this case when the scatterer is at a zero of one of these
states, the square of the wave function of the other state is much smaller there than
its average value.
Many such close zeros should exist to affect
the spectral statistics.
This question is beyond the scope of the present
letter and is left for future research.
Our numerical results are not sensitive enough
to resolve this issue.
The numerical results are used here just to verify that the mean
dependence on the location of the scatterer is given by equation (\ref{totalps}).
We believe that the behavior of the circle billiard
rather than that of the rectangular billiard is typical of integrable
systems perturbed by a localized potential.

In summary, the spectral statistics of the circle billiard, perturbed by
a point-scatterer are intermediate 
between those of the Poisson distribution, characteristic of integrable systems, and of the semi-Poisson distribution. The spectrum 
was shown to be composed of
two uncorrelated components. The first contains energy levels which are nearly unaffected
by the perturbation,
 since the point-scatterer is in a classically forbidden region 
where the wavefunctions
are exponentially small. The relative fraction of
such states was computed analytically and found to depend smoothly  on the location
of the point-scatterer. The second contribution is from states which are affected
by the perturbation. The exact statistics of this level sequence are complicated 
but can be approximated by the semi-Poisson statistics.
The nearest neighbor spacing distribution results of combination of
the two and is a manifestation of the Berry-Robnik statistics.
Other integrable systems should also
exhibit qualitatively similar behavior
 when a localized perturbation is added to them.

\ack
This research was supported in part by the US-Israel Binational
Science
Foundation (BSF) and by the Minerva Center of Nonlinear Physics of Complex Systems.

\end{document}